\documentclass[11pt,showpacs,amsmath,amssymb]{article}
\usepackage[dvips]{graphics}
\usepackage{graphicx}% Include figure files
\usepackage{dcolumn}% Align table columns on decimal point
\usepackage{bm}% bold math
\usepackage{a4wide}

\begin{document}
\begin{center}

{\LARGE \bf On the joint residence time of $N$ independent 
two-dimensional Brownian motions.}

\vspace{0.45in}

{\Large O.B\'enichou$^1$, M.Coppey$^2$, J.Klafter$^3$, M.Moreau$^2$ and G.Oshanin$^2$,}

\vspace{0.45in}

{\sl $^1$ Laboratoire de Physique de la Mati{\`e}re Condens{\'e}e, \\
Coll{\`e}ge de France, 11 Place M.Berthelot, 75252 Paris Cedex 05, France
}

\vspace{0.15in}

{ \sl $^2$ Laboratoire de Physique Th{\'e}orique des Liquides, \\
Universit{\'e} Paris 6, 4 Place Jussieu, 75252 Paris, France}

\vspace{0.15in}

{\sl $^3$ School of Chemistry,\\ Tel Aviv University,  
Tel Aviv 69978, Israel
}

\end{center}

\vspace{0.2in}

\begin{abstract}

We study
 the behavior of 
several joint residence times of $N$ independent
Brownian particles in a 
disc of radius $R$ in two dimensions. 
We consider: (i)
the time $T_N(t)$ spent by all
$N$ particles
simultaneously in the disc 
within the time interval $[0,t]$;
(ii) the time $T_N^{(m)}(t)$ 
which {\it at
least} $m$ out of $N$ particles spend together 
in the disc
within the time interval $[0,t]$;
and (iii) the time ${\tilde T}_N^{(m)}(t)$ 
which {\it exactly} $m$ out of $N$ particles spend together 
in the disc
within the time interval $[0,t]$.
We obtain very simple exact 
expressions for the expectations 
 of
these three residence 
times in the limit $t\to\infty$.
\end{abstract}

\section{Introduction}

One of important and physically
meaningful characteristics of {\it individual}
Brownian particle
trajectories 
is the 
time $T(t)$ spent in a finite domain 
${\cal
S}$ 
within some observation time
$t$. 
This time is referred to  
in the literature as the
"occupation" \cite{Kac,Weisslivre} 
or the "residence time" \cite{Blumen1,Blumen2,Agmon,Zaloj,Bar}.
Note that contrary to the 
first exit  or the first passage times out
of ${\cal S}$, 
the residence time accounts for multiple 
exits from, and entries into ${\cal S}$. As a matter of fact, 
the first passage times can be obtained, once the residence times
are known \cite{Bar}.
The long time properties of  $T(t)$
are 
essentially 
dependent on the dimension of the embedding 
space. In one and
two dimensions 
all moments of $T(t)$ diverge with time, 
which
mirrors the fact that here 
the Brownian motion is "compactly" exploring the space \cite{pgg}.
In  two dimensions, in particular, the long time
behavior of $T(t)$ obeys the  Kallianpur-Robbins' law
\cite{Kallianpur}, which states that the scaling 
variable $T'(t)$, defined as 
$\displaystyle T'(t)=\frac{4\pi  D T(t)}{S \ln{t}}$, 
where $D$ stands for
the particle's diffusion  coefficient and $S$ is the area
of the domain, 
is asymptotically
distributed according to the exponential law. On the contrary, in
higher dimensions, these moments tend to
finite limiting values
when $t\to\infty$. 
Unfortunately, explicit 
expressions in these
transient cases are not always
available. A notable exception
is the special case of a three-dimensional spherical domain, a case 
in which
explicit values of 
the moments of the
limiting values of the occupation time
have been obtained 
\cite{Zaloj}.

Recently, motivated in part by 
the development of new experimental techniques,
such as, e.g.,  
fluorescence correlation spectroscopy (FCS), which enables to
register 
single particle 
events in many 
particle systems (see for example
Ref.\cite{Eigen}),
there has been a considerable 
interest in the behavior of 
related properties of  
{\it a set of independently diffusing
particles} (see Ref.\cite{Acedo} and references therein). 
For instance, 
the
number of distinct sites visited in $n$ steps 
by $N$ random walks on
a lattice  \cite{Larralde}, the time
spent together by the first $j$ out of a total number $N$ 
particles, before they escape
from a given region 
\cite{Weiss,Yuste1,Klafter2} or the order
statistics  
for $N$ diffusing particles in 
the trapping problem
\cite{Acedo} have been extensively analysed. 
It has been
recognized that, 
despite the absence of any physical
interaction between the diffusing particles, 
in many instances a 
highly 
cooperative behavior
emerges. 
%Most of these studies have focussed 
%only on the large $N$ asymptotic limit. 

In this paper 
we study collective statistical
properties of the {\it mean}
residence time 
of $N$ independent 
Brownian particles
in a finite domain
${\cal S}$. We focus here on the behavior in 
a two-dimensional
continuum and 
suppose that  ${\cal S}$ is a 
disc of radius $R$ centered at the origin. We
consider 
three kinds of occupation times:
\begin{itemize}
\item The time $T_N(t)$ spent by all
$N$ particles
simultaneously in the domain  ${\cal S}$ 
within the time interval $[0,t]$;
\item The time $T_N^{(m)}(t)$ 
which {\it at
least} $m$ out of $N$ particles spend together 
in the domain ${\cal S}$
within the time interval $[0,t]$;
\item The time ${\tilde T}_N^{(m)}(t)$ 
which {\it exactly} $m$ out of $N$ particles spend together 
in the domain ${\cal S}$
within the time interval $[0,t]$.
\end{itemize}
We obtain here very simple expressions for the mean values  of
these times in the limit $t\to\infty$.

We note that this type of functionals of random walks 
has not 
received much
attention up to now, except for two particular cases. 
The first one  concerns 
the occupation time
of harmonically bounded Brownian particles in two
dimensions \cite{ol}.
In the second case the problem of the 
occupancy of a single lattice
 site by a {\it concentration} of 
random walkers has been analysed \cite{boguna}. 
Note that in both cases one finds
that 
the moments of the residence time
diverge as $t \to \infty$.

The analysis of 
such functionals of Brownian trajectories
might be of importance for several 
physical processes. Consider
for example $N$ molecules diffusing on a surface, which contains a
receptor sensible to the presence of a given number of molecules,
say $m$, with $m\le N$, i.e. we suppose that there is a kind of
sensitivity threshold. Suppose next that the activity of the
receptor is proportional to the time during which it is active,
that is to the time when there are at least $m$ molecules in the
vicinity of the receptor. We finally have a situation where the
response of the receptor at time $t$ is proportional to the time
$T_N^{(m)}$ defined previously. Another application can be found
in the FCS which is, as mentioned, a single
molecule microscopy method. In these experiments, a given region
of space is illuminated by a laser beam, and one observes the
fluorescence signal of the molecules going through that region. In
order to observe a single molecule, it might be  important to
limit events corresponding to the presence of more than one
molecule  inside the illuminated region, by decreasing the
extension of the beam \cite{Eigen}. The evaluation of the 
time spent when there
are at least 2 particles out of the total number of molecules
inside the illuminated region, that is precisely the quantity
$T_N^{(2)}$, could give a quantitative measurement of these
undesirable events, and thus could furnish an indication of the
required extension of the beam.

\section{The model.}

Consider $N$ independent Brownian particles diffusing on an infinite
two-dimensional plane. Let ${\bf r}_j(t)$ denote 
the positions of these
particles at time $t$, while 
$D_j$ 
stand for
the corresponding diffusion coefficients, 
which are not necessarily equal to each other. 
We also assume that 
all particles are initially at the origin, i.e. ${\bf r}_j^{(0)} =0 $. 

Next, let us introduce three auxiliary indicator functions
${\bf
1}_{\cal S}\left({\bf r}\right)$, $I_{\cal S}^{(m)}\left(\{{\bf r}_j\}\right)$ and  
${\tilde I}_{\cal S}^{(m)}\left(\{{\bf r}_j\}\right)$
which have the following properties
\begin{eqnarray}
{\bf 1}_{\cal S}({\bf r})=
\cases{1,&if $|{\bf r}| \leq R,$\cr0,&otherwise,\cr}\nonumber\\
\end{eqnarray}
\begin{eqnarray}
I_{\cal S}^{(m)}\left(\{{\bf r}_j\}\right) =
\cases{1,&if at least $m$ of $N$ particles are in ${\cal S},$ \cr0,&otherwise,\cr}\nonumber\\
\end{eqnarray}
and
\begin{eqnarray}
{\tilde I}_{\cal S}^{(m)}\left(\{{\bf r}_j\}\right)=
\cases{1,&if exactly $m$ of $N$ particles are in ${\cal S},$ \cr0,&otherwise.\cr}\nonumber\\
\end{eqnarray}
The occupation times $T_N(t)$, $T_N^{(m)}(t)$ and ${\tilde T}_N^{(m)}(t)$, 
defined in the Introduction, 
can be then formally written
as
\begin{equation}\label{tempsavectoutes}
T_N(t)=\int _0^t{\rm d}t'\;\left(\prod_{j=1}^N{\bf 1}_{{\cal
S}}({\bf r}_j(t'))\right),
\end{equation}
\begin{equation}
T_N^{(m)}(t)= \int_0^t{\rm d}t'\; I_{\cal S}^{(m)}\left(\{{\bf r}_j(t')\}\right),
\end{equation}
and
\begin{equation}
{\tilde T}_N^{(m)}(t)=\int_0^t{\rm
d}t'\; {\tilde I}_{\cal S}^{(m)}\left(\{{\bf r}_j(t')\}\right)
\end{equation}
Using the Poincare-type 
formulae \cite{feller}, the two last equations can 
be cast into a more convenient
form:
\begin{equation}\label{tempsaumoins}
T_N^{(m)}(t)= \int_0^t{\rm d}t'\;\sum_{k=m}^N(-1)^{(k-m)}{k-1
\choose m-1}\sum_{1\le i_1<...<i_k\le N} \left(\prod_{j=1}^k{\bf
1}_{{\cal S}}({\bf r}_{i_j}(t'))\right)\
,
\end{equation}
\begin{equation}\label{tempsexactement}
{\tilde T}_N^{(m)}(t)=\int_0^t{\rm
d}t'\;\sum_{k=m}^N(-1)^{(k-m)}{k \choose m}\sum_{1\le
i_1<...<i_k\le N} \left(\prod_{j=1}^k{\bf 1}_{{\cal S}}({\bf
r}_{i_j}(t')) \right)\,,
\end{equation}
where the sums extend over all ordered $k$-uplets in
$\{1, \ldots , N\}$, and $N \geq m \geq 2$. Eqs.(\ref{tempsaumoins})
and (\ref{tempsexactement})
show
that $T_N^{(m)}(t)$ and ${\tilde
T}_N^{(m)}(t)$ are both functionals of $T_N(t)$. Consequently, 
we focus our analysis on the behavior 
of $\langle T_N(t) \rangle$. Results 
for $\langle T_N^{(m)}(t) \rangle$ and $\langle {\tilde T}_N^{(m)}(t) \rangle$ 
can be straightforwardly obtained once $\langle T_N(t) \rangle$ is known and
will be presented here without derivation.

\section{Results.}

The first moment of
$T_N(t)$, i.e. the mean time spent in ${\cal S}$ by $N$ particles
simultaneously within the time interval $[0,t]$, is given by
\begin{equation}\label{moment}
\mu_{N}(t)=\langle T_N(t) \rangle 
=\int_0^t \left(\prod_{j=1}^N\langle {\bf
1}_{{\cal S}}({\bf r}_j(t'))\rangle _j\right){\rm d}t'.
\end{equation}

Turning to the limit $t \to \infty$, and
performing averaging over the realizations of Brownian motions,
we obtain 
\begin{equation}
\label{integral}
 \mu_{N} \equiv \lim_{t = \infty} \mu_N(t) =  \int_0^{\infty} {\rm d}t'
\left\{\prod_{j=1}^N\left[1-\exp\left(-\frac{R^2}{4t'D_j}\right)\right]\right\}
\end{equation}
We take next the 
diffusion coefficient $D_1$
of the first particle as the reference. Introducing the
corresponding time
scale $\displaystyle \tau = R^2/4D_1$, 
as well as the dimensionless
parameters 
$\displaystyle \lambda_i=\frac{D_1}{D_i}$, $i = 1, \ldots, N$, we find 
that $\mu_{N}$ obeys
\begin{equation}
\mu_{N}= \tau \int_0^1
\;\frac{\prod_{j=1}^N\left(1-v^{\lambda_j}\right)}{v(\ln
v)^2}\;{\rm d}v
.
\end{equation}
Next, expanding the 
product 
$\prod_{j=1}^N\left(1-v^{\lambda_j}\right)$ in powers of $v$, we 
arrive, by performing integrations by parts, at the following explicit expression:
\begin{equation}\label{mu}
\mu_{N}=\tau \sum_{ k=1}^N\;(-1)^k \sum_{1\le i_1<i_2<...<i_k\le
N}\left(\lambda_{i_1}+...+\lambda_{i_k}\right)\ln\left(\lambda_{i_1}+...+\lambda_{i_k}\right).
\end{equation}
which does not seem to be an {\it
a priori} trivial result. 

In the particular case 
 when only two Brownian particles, $N=2$, are present in the system, 
Eq.(\ref{mu}) yields the following symmetric compact expression:
\begin{equation}\label{N=2}
\mu_{2}=\frac{R^2}{4}\left[\frac{1}{D_1}\ln\left(1+\frac{D_1}{D_2}\right)+\frac{1}{D_2}\ln\left(1+\frac{D_2}{D_1}\right)\right]
\end{equation}
When one of the particles moves
much faster than the second one, 
that is, for instance, when $D_1\gg D_2$,
Eq.(\ref{N=2}) reduces to
\begin{equation}\label{limiteD1D2}
\mu_{2}\sim \frac{R^2}{4D_1}\ln\left(\frac{D_1}{D_2}\right).
\end{equation}
Note that such a result appears to be quite reasonable  
from the physical point of view. 
To a  first approximation, one estimates that 
the
joint residence time 
is given by the joint residence time before the
first exit time of the slow particle out of the disc. This time is of
order $\displaystyle t_2=R^2/D_2$. Meanwhile, the fast particle
leaves the disc and returns 
back several times. The time spent at each
return of the fast particle inside the disc 
is of order $\displaystyle
\tau=R^2/D_1$, and the order of magnitude of the number of
returns at time $t_2$ of the particle one inside the cercle is given by
$\displaystyle \ln\left(t_2/\tau\right)$ (cf. the Kallianpur-Robbins' law 
mentioned in the Introduction). As a consequence, we
expect that for $D_1\gg D_2$ one has
\begin{equation}
\mu_{2}\sim
\tau \ln\left(\frac{t_2}{\tau}\right)\sim
\frac{R^2}{D_1}\ln\left(\frac{D_1}{D_2}\right),
\end{equation}
which reproduces the dependence  of $\mu_{2}$ on the
diffusion coefficients $D_1$ and $D_2$ given by Eq.
(\ref{limiteD1D2}).

Consider now the special case when all of
the particles have the same diffusion 
coefficient $D_1$. Here, Eq.
(\ref{mu}) becomes
\begin{equation}\label{memeD}
\mu_{N}=\tau \;\sum_{k=1}^N\;(-1)^k k {N \choose k} \ln(k),
\end{equation}
which in the asymptotical limit 
$N\to\infty$ behaves as
\begin{equation}\label{limitegrandN}
\mu_{N}\sim  \frac{\tau}{\ln(N)}
,
\end{equation}
displaying 
a very slow decay as a 
function of $N$. 

As a matter of fact,
this expression coincides 
exactly 
with the result of Refs.\cite{kat,kat1} obtained for
the mean first exit time out of
the disc 
by one of $N$ particles. This is, of course, not counterintuitive,
since, after
the departure 
of one particle, the probability that all
$N$ particles arrive altogether inside the disc 
goes
to zero when $N$ tends to infinity.

Note also that the result
in Eq.(\ref{limitegrandN}) is completely 
different from its lattice counterpart, e.g., from the result obtained for
the mean joint occupation
time of a single lattice site by $N$ independent random walks. 
Here, for example,  for continuous
time random walks (with jump
frequency $\omega$) on a $d$-dimensional 
hypercubic lattice, 
one finds (see Ref.\cite{boguna}):
\begin{eqnarray}
\mu_{N}=\int_0^\infty\prod_{j=1}^N p_j({\bf 0},t'|{\bf 0},0){\rm
d}t' =\int_0^\infty e^{-N \omega t'}\left[{\rm
I}_0\left(\frac{\omega t'}{d}\right)\right]^{Nd} dt',
\end{eqnarray}
where $p_j({\bf 0},t'|{\bf 0},0)$ denotes 
the probability 
that the  $j$ walker, which starts 
its random walk at the origin, returns to the origin
at time $t'$,
and ${\rm I}_0(z)$ is the modified Bessel function.
The large-$N$ asymptotic behavior of $\mu_N$ thus follows
\cite{Hughes}
\begin{equation}
\mu_{N}=\frac{1}{N\omega}\left(1+\frac{1}{Nd}+\frac{3}{4(Nd)^2}+\frac{3}{2(Nd)^3}+...\right)
\end{equation}
In this case, $\mu_{N}$ decreases with the increase in the number
of walkers $N$ at a  much faster rate than in continuum, Eq.(\ref{limitegrandN}). 
A similar result 
has been already reported in the context
of the behavior of
first passage times in finite systems \cite{Weiss}.

We 
finally consider an important special case
in which only one of the particles has the
diffusion coefficient $D_1$, while
the remaining $N-1$ 
have the diffusion coefficient $D_2$ - a case
of an "impure" particle  among a 
set of  $N-1$ "pure" particles. In this special case
Eq.(\ref{mu}) becomes, explicitly, 
\begin{eqnarray}
\mu_{N}&=&\frac{R^2}{4}\left\{\frac{N-1}{D_2}\sum_{k=1}^{N-1}(-1)^k\ln\left(\frac{k}{k+D_2/D_1}\right){N-2
 \choose k-1}\right.\nonumber\\
 &+&\left.\frac{1}{D_1}\sum_{k=0}^{N-1}(-1)^{k-1}\ln\left(k+D_2/D_1\right){N-1
 \choose k}\right\}.
\end{eqnarray}

\vspace{1cm}

We turn now to the behavior of $\displaystyle
\mu_{N}^{(m)}\equiv\lim_{t\to\infty}\langle
T_N^{(m)}(t)\rangle$. Assuming
 that all  particles have the same diffusion
coefficient $D_1$, we find, 
using 
Eq.(\ref{tempsaumoins}), as well as the result obtained 
previously
for $\mu_{N}$, Eq.(\ref{memeD}), that
\begin{eqnarray}
  \mu_{N}^{(m)}=\tau \sum_{k=m}^N(-1)^{(k-m)}{k-1 \choose m-1}{N
  \choose k} \sum_{j=1}^k(-1)^j j {k \choose j}\ln(j).
\end{eqnarray}
Changing the order of summations, we finally obtain
the following result:
\begin{eqnarray}
  \mu_{N}^{(m)}=\tau m {N \choose m}
  \sum_{k=0}^{m-1}(-1)^{m-k}{m-1 \choose k} \ln (N-k).
\end{eqnarray}
In particular, for $m=2$ (which corresponds to the FCS example
mentioned in the Introduction), the last equation yields
\begin{eqnarray}\label{aumoinsresultat}
  \mu_{N}^{(2)}=\tau N (N-1) \ln\left(\frac{N}{N-1}\right),
\end{eqnarray}
where $\tau = R^2/D_1$. 
The asymptotic limit  $N\to\infty$ with $m$ finite in
Eq.(22) leads to
\begin{eqnarray}
  \mu_{N}^{(m)}\sim\frac{N}{m-1} \tau,
\end{eqnarray}
which is compatible, for  $m = 2$, with the result in Eq.(23)
when $N \to \infty$.

Lastly, for  $\displaystyle {\tilde
\mu}_{N}^{(m)}\equiv\lim_{t\to\infty}\langle {\tilde
T}_N^{(m)}(t)\rangle$, i.e. the mean limiting time spent simultaneously by exactly $m$
out of $N$ particles within ${\cal S}$, we find, using Eqs.(\ref{tempsexactement}) 
and (\ref{memeD}) (or the relation ${\tilde
\mu}_{N}^{(m)}=\mu_{N}^{(m)}- \mu_{N}^{(m+1)})$, 
\begin{eqnarray}\label{exactementresultat}
 {\tilde
\mu}_{N}^{(m)} =\tau {N \choose m} \sum_{k=0}^m(-1)^{m-k}(N-k){m
\choose k} \ln(N-k),
\end{eqnarray}
We note parenthetically that, curiously enough, ${\tilde
\mu}_{N}^{(m)} $ appears to be a non-monotonic function of $m$, as suggested by numerical analysis
of Eq.(\ref{exactementresultat}).
In the asymptotic limit 
$N\to\infty$ with $m$ finite, 
Eq.(\ref{exactementresultat}) yields
\begin{eqnarray}
  {\tilde \mu}_{N}^{(m)}\sim\frac{N}{m(m-1)} \tau,
\end{eqnarray}
which is not an $\it apriori$ trivial result.

\section{Conclusions.}

To conclude, we have studied
several types of joint residence times
of a disc of radius $R$ 
by $N$ independent Brownian particles:
The time $T_N(t)$ spent by all
$N$ particles
simultaneously in the domain  ${\cal S}$ 
within the time interval $[0,t]$;
the time $T_N^{(m)}(t)$ 
which {\it at
least} $m$ out of $N$ particles spend together 
in the domain ${\cal S}$
within the time interval $[0,t]$;
and finally, the time ${\tilde T}_N^{(m)}(t)$ 
which {\it exactly} $m$ out of $N$ particles spend together 
in the domain ${\cal S}$
within the time interval $[0,t]$.
We have shown that in case when all the particles are
initially located 
at the center of the disc, 
it is possible to obtain
the mean values of such joint residence times 
exactly for arbitrary values of the particle number $N$.

\bibliography{biblio}

\end{document}